\documentclass{article}
\usepackage{styles/spconf}
\usepackage{styles/defs}

\usepackage{amsmath,graphicx}
\usepackage{xcolor}
\hypersetup{hidelinks}

\title{Learning Array Signal Topologies as Conditional Neural Manifolds}

\name{%
  Julian P. Merkofer$^{\star, \dagger}$%
  \qquad  Vincent van de Schaft$^{\star}$%
  \qquad Ruud J. G. van Sloun$^{\star}$%
}
\address{%
  $^{\star}$ Eindhoven University of Technology, Eindhoven, The Netherlands \\%
  $^{\dagger}$ University Medical Center Utrecht, Utrecht, The Netherlands%
}

\begin{document}
\ninept
\setlength{\abovedisplayskip}{4pt plus 1pt minus 2pt}
\setlength{\belowdisplayskip}{4pt plus 1pt minus 2pt}
\setlength{\abovedisplayshortskip}{2pt plus 1pt}
\setlength{\belowdisplayshortskip}{2pt plus 1pt}

\maketitle

\begin{abstract}
\vspace{-1mm}
Subspace methods such as \ac{music} achieve super-resolution \ac{doa} estimation by exploiting the orthogonality between the array manifold and the noise subspace of the measurements. Their accuracy therefore depends on the assumed manifold and degrades under model mismatch, while parameters not identifiable from the spatial manifold cannot be recovered. In this work, we propose the \ac{cnm}, which replaces the fixed manifold with an observation-conditioned mapping from source parameters to steering vectors. An encoder maps the snapshots to a latent scene representation that conditions a zero-initialized neural field over the parameter space. The manifold is learned without steering-vector supervision by shaping the resulting \ac{music} landscape. Since the correction acts on the manifold rather than on the estimator, it can be used by other manifold-based methods without modification. The \ac{cnm} restores resolution under array imperfections, colored noise, correlated sources, and near-field propagation, and resolves the angle-frequency ambiguity inherent to the nominal spatial manifold.
\end{abstract}
%

\vspace{-1mm}
\section{Introduction} \label{sec:intro}
\vspace{-2mm}

\acresetall

Estimating the parameters of wave sources from the snapshots of a sensor array is a core problem of array signal processing, with applications ranging from radar and communications to medical imaging \cite{krim1996twodecades, viberg1997twodecades, liu2023sam, pesavento2023threedecades}. Subspace methods, most prominently \ac{music} \cite{Schmidt1986MUSIC} and its variants \cite{barabell1983rootmusic, shan1985smoothing, roy1989esprit}, achieve resolution beyond the beamwidth of the aperture by exploiting the orthogonality between the source steering vectors and the noise subspace of the covariance matrix. Their accuracy rests on the array manifold, the analytical map from the source parameters to the array response, which is assumed to be known exactly.

In practice, both the manifold and the signal model can deviate from these assumptions.
Gain and phase errors, displaced elements, and mutual coupling change the array response \cite{swindlehurst1992modelerrors, friedlander1991coupling}, near-field propagation curves the wavefront \cite{huang1991nearfield}, and broadband sources spread the response over frequency \cite{wang1985coherent}, while correlated sources \cite{shan1985smoothing} and spatially colored noise alter the subspaces estimated from the measurements.
Classical methods address these limitations separately by calibrating the array response \cite{friedlander1991coupling}, modeling it in a parametric basis \cite{belloni2007manifold}, bounding its mismatch in robust beamforming \cite{vorobyov2003robust}, extending the signal model to the near field \cite{huang1991nearfield} or to broadband sources \cite{wang1985coherent}, or preprocessing the covariance through spatial smoothing \cite{shan1985smoothing}. 

More recently, data-driven estimators learn a direct mapping from the measurements to the source parameters \cite{liu2018imperfections, papageorgiou2021gridcnn}, while model-based deep-learning methods \cite{shlezinger2023modelbased} retain the structure of subspace estimation and learn the covariance, the subspaces, or the read-out of the resulting spectrum \cite{barthelme2021reconstruction, shmuel2023deeprootmusic, shmuel2025subspacenet, merkofer2022damusicicassp, merkofer2024damusic}. In all of these methods, the learning acts on the statistics that enter the estimator or on the read-out of its output, while the array manifold itself remains the nominal one, or is discarded altogether.

In this work, we propose to learn the array manifold itself through a \ac{cnm}. The contribution is threefold. First, the \ac{cnm} replaces the fixed manifold with an observation-conditioned mapping from source parameters to steering vectors, allowing the manifold to adapt to the measured scene. Second, conditioning on the complete snapshots allows the learned manifold to not only restore consistency with the measured subspace when the nominal assumptions fail, but also to extend the identifiable parameter space using information that is absent from the spatial response. Third, the manifold is learned without supervision of the true steering vectors by optimizing the resulting \ac{music} spectrum toward a target determined by the spatial and temporal distinguishability of the source parameters.
We demonstrate the proposed approach under array imperfections, colored noise, correlated sources, near-field propagation, and joint angle--carrier estimation.\footnote{The implementation code can be found online at: \url{https://github.com/julianmer/CNM-MUSIC}.}

\vspace{-1mm}
\section{Method} \label{sec:method}
\vspace{-2mm}

\begin{figure*}
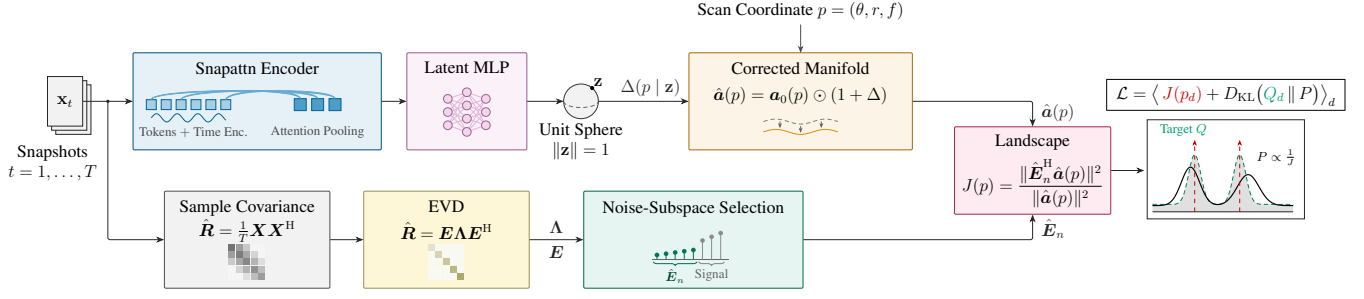

\centering
\vspace{-1mm}
\includestandalone[width=\textwidth]{figures/method}
\vspace{-4mm}
\caption{The \ac{cnm} pipeline. Snapshots are encoded into a latent scene representation $\vect{z}$ that conditions a zero-initialized neural field deforming the nominal manifold as $\hat{\vect{a}}(p \mid \vect{z})=\vect{a}_0(p)\odot(1+\Delta(p\mid\vect{z}))$. The corrected manifold is evaluated with the unchanged \ac{music} subspace estimation, while $P\propto1/J$ is shaped toward a target $Q$ with one peak per source.\vspace{0mm}}
\label{fig:method}
\end{figure*}

\vspace{-1mm}
\subsection{Problem Formulation}
\vspace{-1mm}

We consider the estimation of source parameters $p_d \in \mathcal{P}$ from $D$ signals measured by an array of $M$ sensors. In its general form, the signal received at sensor $m$ is
\begin{equation} \label{eq:model}
x_m(t) =
\sum_{d=1}^{D}
g_m(p_d)\,
s_d\left(t-\tau_m(p_d)\right)
+
v_m(t),
\end{equation}
where $s_d(t)$ is the $d$-th source signal, $\tau_m(p_d)$ its propagation delay relative to a reference sensor, $g_m(p_d)$ the corresponding gain, and $v_m(t)$ additive noise. The measurements across the array form the snapshots
$\boldsymbol{x}_t=[x_1(t),\dots,x_M(t)]^{\mathrm T}$ and
$\boldsymbol{X}=[\boldsymbol{x}_1,\dots,\boldsymbol{x}_T]$.

For a frequency component $f$, the corresponding array response is described by the steering vector
\begin{equation} \label{eq:steer}
[\boldsymbol{a}(p)]_m
=
g_m(p) e^{-j2\pi f\tau_m(p)},
\qquad
p=(\theta,r,f,\ldots),
\end{equation}
where the coordinates of $p$ depend on the estimation problem. This formulation includes arbitrary array geometries and both far- and near-field propagation, while broadband signals are represented by their frequency components.

For example, the nominal far-field response of a \ac{ula} with half-wavelength spacing at the design frequency $f_c$ is
\begin{equation} \label{eq:ula}
[\boldsymbol{a}_0(\theta,f)]_m
=
e^{-j\pi (m-1) (f / f_c) \sin\theta}
\xrightarrow{f=f_c}
e^{-j\pi (m-1)\sin\theta},
\end{equation}
where the latter is the conventional narrowband manifold. In the near field, the plane-wave delay is replaced by the exact path difference determined by $(\theta,r)$.

\noindent
\Ac{music} \cite{Schmidt1986MUSIC} decomposes the sample covariance matrix
into signal and noise subspaces, with the $M-D$ eigenvectors corresponding to the smallest eigenvalues forming the estimated noise subspace $\hat{\boldsymbol{E}}_n$. A candidate steering vector is then evaluated by its projection onto this subspace,
\vspace{-4mm}
\begin{equation} \label{eq:null}
J(p)
=
\frac{\|\hat{\boldsymbol{E}}_n^{\mathrm H}\boldsymbol{a}(p)\|^2}
{\|\boldsymbol{a}(p)\|^2}.
\end{equation}
At the true source parameters, $\boldsymbol{a}(p_d)$ lies in the signal subspace and is therefore orthogonal to $\hat{\boldsymbol{E}}_n$, yielding minima of $J(p)$ and corresponding peaks of $1/J(p)$.

\vspace{-1mm}
\subsection{Conditional Neural Manifold}
\vspace{-1mm}

\begin{figure*}
    \centering
    \vspace{-1mm}
    \includegraphics[width=0.875\textwidth]{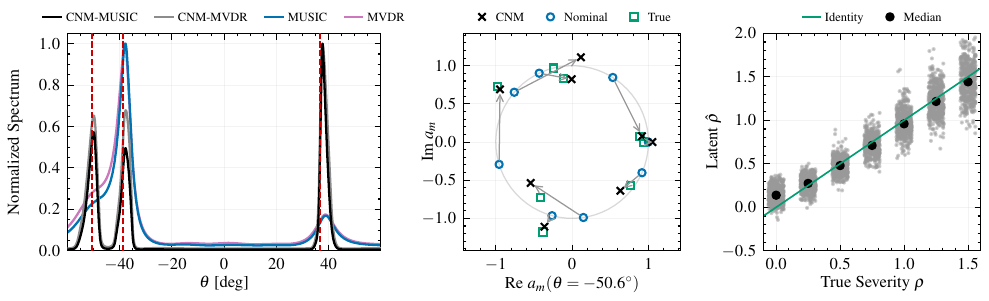}
    \vspace{-4mm}
    \caption{Array imperfections of severity $\rho = 1$ at 15 dB, $T = 200$, and $D = 3$. Left: peak-normalized spectra of \ac{music} and \ac{mvdr} on the nominal and on the corrected manifold, true \acp{doa} dashed. Center: the steering vector at one true \ac{doa} with arrows marking the correction. Right: linear read-out of the severity $\rho$ from the latent $\vect{z}$ against the truth, cross-validated over 3500 scenes ($R^2 = 0.91$).\vspace{-1mm}}
    \label{fig:qualitative}
\end{figure*}

\begin{figure*}
    \centering
    \vspace{-1mm}
    \begin{minipage}[t]{0.700\textwidth}
        \centering
        \includegraphics[width=0.95\textwidth]{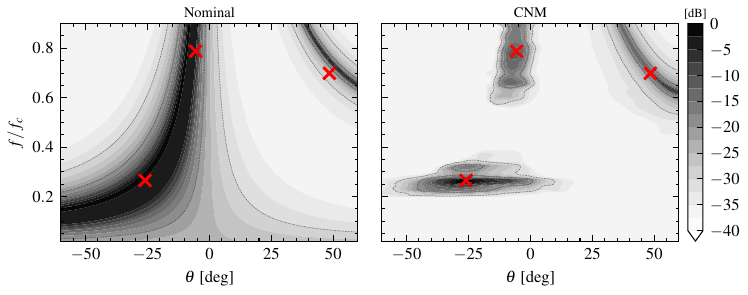}%
        \vspace{-4mm}
        \caption{Joint angle-carrier estimation of $D = 3$ \ac{ofdm} sources at independent carriers (15 dB, $T = 200$). Left: the null spectrum over $(\theta, f)$ on the nominal manifold. Right: the corrected landscape with isolated peaks at the true parameters (crosses).}
        \label{fig:joint_carrier}
    \end{minipage}\hfill%
    \begin{minipage}[t]{0.266\textwidth}
        \centering
        \includegraphics[width=0.975\textwidth]{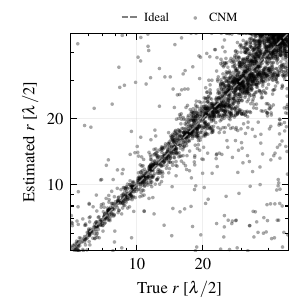}%
        \vspace{-4mm}
        \caption{Estimated versus true near-field range over 3000 sources spanning the full range (15 dB, $T = 200$).}
        \label{fig:readouts}
    \end{minipage}
    \vspace{-1mm}
\end{figure*}

A \ac{cnm} replaces the fixed nominal manifold with an observation-conditioned one (Fig.~\ref{fig:method}),
\begin{equation} \label{eq:cnm}
\hat{\boldsymbol{a}}(p \mid \boldsymbol{z})
=
\boldsymbol{a}_0(p)
\odot
\bigl(1+\Delta(p \mid \boldsymbol{z})\bigr),
\qquad
\boldsymbol{z}=h(\boldsymbol{X}),
\end{equation}
where the encoder $h$ maps the snapshots to a latent representation $\boldsymbol{z}$ and $\Delta(\cdot\mid\boldsymbol{z})$ is a neural field \cite{xie2022neuralfields} over the source-parameter space. The conventional mapping $p\mapsto\boldsymbol{a}_0(p)$ therefore becomes conditioned on the observed signals through $\boldsymbol{z}$.
The \ac{cnm} therefore adapts the manifold to the observed scene while preserving the structure of the downstream estimator.
The encoder operates directly on the snapshots $\boldsymbol{X}$. Each snapshot is represented by the real and imaginary parts of $\boldsymbol{x}_t$ and of the upper triangle of $\boldsymbol{x}_t\boldsymbol{x}_t^{\mathrm H}$, together with a sinusoidal encoding of $t$. The resulting tokens are pooled by cross-attention onto $K=8$ learned queries \cite{lee2019settransformer} and mapped to $\boldsymbol{z}\in\mathbb{R}^{32}$, which is normalized to the unit sphere \cite{yue2026sphereencoder}. The neural field encodes the scan coordinate $p$ using fixed harmonic features, concatenates these with $\boldsymbol{z}$, and maps them through an \ac{mlp} with two hidden layers of 128 units to the real and imaginary parts of $\Delta$.
The final layer of $\Delta$ is initialized to zero, such that $\Delta\equiv0$ before training and
$\hat{\boldsymbol{a}}(p\mid\boldsymbol{z})=\boldsymbol{a}_0(p)$. The untrained \ac{cnm} therefore exactly recovers the nominal manifold.
Estimation proceeds by replacing the nominal steering vector by the learned one. 
The spectrum is evaluated over the scan coordinates in $\mathcal{P}$ and the $D$ largest peaks of $1/J$ give the source estimates. Since the learned quantity is the manifold itself, the same $\hat{\boldsymbol{a}}(p\mid\boldsymbol{z})$ can be used directly by other manifold-based estimators, such as \ac{mvdr}.

\vspace{-1mm}
\subsection{Training}
\vspace{-1mm}

The \ac{cnm} is trained on simulated scenes with known source parameters $\{p_d\}_{d=1}^{D}$. Rather than supervising the learned manifold against the true array response, we optimize it through the resulting \ac{music} spectrum. The loss combines a point-wise subspace constraint with a distributional constraint on the complete spectrum,
\begin{equation} \label{eq:loss}
    \mathcal{L}
    =
    \underbrace{
        \left\langle J(p_d) \right\rangle_d
    }_{\text{subspace consistency}}
    +
    \underbrace{
        \left\langle D_{\mathrm{KL}}\bigl(Q_d(p) \,\Vert\, P(p)\bigr) \right\rangle_d
    }_{\text{spectrum shaping}} .
\end{equation}
Here, $p$ denotes the scan coordinate and $p_d$ the true parameter of source $d$. The spectrum $P(p)\propto 1/J(p)$ is compared to a target $Q_d(p)$ centered at $p_d$. The first term directly enforces a null at the true source parameter by minimizing the projection of the learned steering vector onto the estimated noise subspace. The second term shapes the complete spectrum around each source.

The target is defined as
\begin{equation} \label{eq:target}
    Q_d(p)
    \propto
    e^{-\tilde{\delta}_d(p)/\varepsilon},
\end{equation}
where $\tilde{\delta}_d(p)$ describes the distinguishability between a candidate parameter $p$ and the true source parameter $p_d$. We define
\begin{equation} \label{eq:deficit}
    \tilde{\delta}_d(p)
    =
    1
    -
    \frac{
        \left|
            \boldsymbol{a}_0(p_d)^{\mathrm{H}}
            \boldsymbol{a}_0(p)
        \right|^2
    }{
        \|\boldsymbol{a}_0(p_d)\|^2
        \|\boldsymbol{a}_0(p)\|^2
    }
    \left|
        \frac{D_T(f-f_d)}{T}
    \right|^2,
\end{equation}
with $D_T$ denoting the Dirichlet kernel of the $T$-sample record. The steering-vector overlap measures the similarity of the spatial array responses, while the Dirichlet term accounts for the frequency resolution of the temporal record. The target is therefore determined by the distinguishability provided jointly by the array and the record rather than by a fixed distance in parameter space. When frequency is fixed, the Dirichlet term equals one and $\tilde{\delta}_d(p)$ reduces to the normalized overlap deficit between the nominal steering vectors.

The scale $\varepsilon$ controls the width of the target and is set from the minimum source separation of the training data. For each scene, $P(p)$ and $Q_d(p)$ are evaluated and normalized over $N=2048$ parameters sampled uniformly from $\mathcal{P}$ before evaluating the \ac{kl} divergence. The noise subspace is obtained from the sample covariance exactly as during inference.

\vspace{-1mm}
\section{Experiments} \label{sec:experiments}
\vspace{-2mm}

\begin{figure*}
    \centering
    \vspace{-1mm}
    \includegraphics[width=0.975\textwidth]{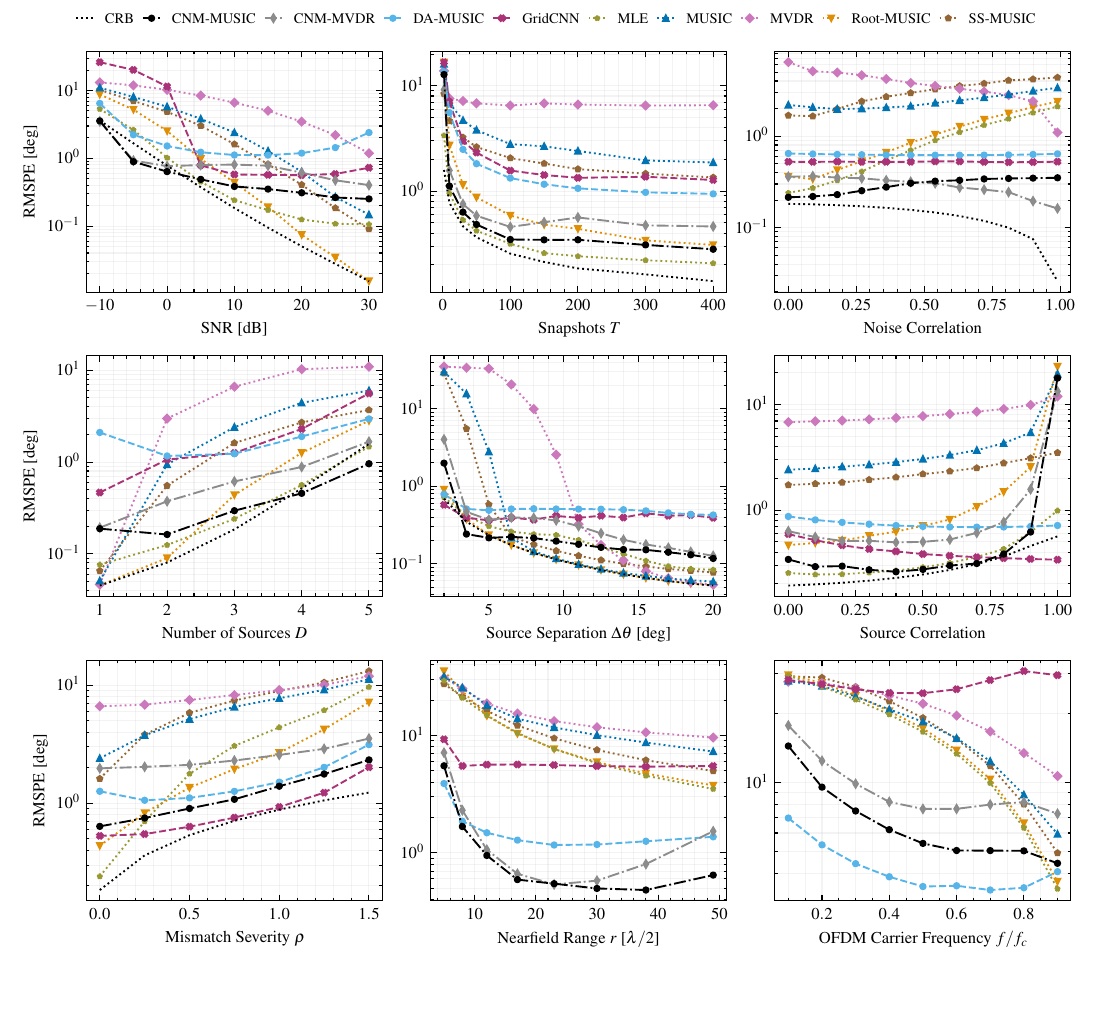}
    \vspace{-12mm}
    \caption{\ac{rmspe} versus the varied condition, all other conditions at their default values, $10^4$ Monte-Carlo scenes per point. Top row: \ac{snr}, number of snapshots, and noise correlation; middle row: number of sources, separation of two sources, and source correlation; bottom row: severity of the array imperfections, near-field range, and \ac{ofdm} carrier frequency. The \ac{crb} is evaluated with the true manifold.
    }
    \label{fig:overview}
\end{figure*}

\vspace{-1mm}
\subsection{Setup}
\vspace{-1mm}
All scenes are simulated from \eqref{eq:model} for a \ac{ula} with $M = 8$ half-wavelength spaced elements and a field of view of $\pm 60^\circ$, scanned on a shared grid of 480 points. The source signals and the noise are drawn from the complex Gaussian distribution, normalized to meet the \ac{snr} per source, and the \acp{doa} are drawn uniformly at least $2^\circ$ apart. Unless stated otherwise, the \ac{snr} is 10 dB, $T = 200$, $D = 3$, the sources are uncorrelated, the noise is white, and the array is calibrated and in the far field.

Each scenario varies exactly one of these conditions over a range, and a separate model is trained per scenario: the \ac{snr} in $[-10, 30]$ dB; $T \in [1, 400]$; $D \in [1, 5]$; two sources with separation $\Delta\theta$; source correlation in $[0, 1]$; spatially colored noise with covariance $c^{|i - j|}$, $c \in [0, 0.99]$; array imperfections of severity $\rho \in [0, 1.5]$, which scales gain ($\pm 20\%$), phase ($\pm 30^\circ$), and position ($\pm 0.2$ spacings) errors drawn uniformly per element and scene, and a Toeplitz mutual coupling of strength $0.3 e^{j\pi/3}$ \cite{liu2018imperfections}; near-field sources at $r \in [5, 49]$ half-wavelengths; and \ac{ofdm} sources of 10 subcarriers spanning 10\% of the band at per-source carriers $f / f_c \in [0, 0.9]$.

Where the range or the carrier is free, one \ac{cnm} is trained with it pinned at its nominal value, so that the correction absorbs it like any other deformation, and one with it active, which estimates it jointly with the \acp{doa}. The estimators are evaluated on $10^4$ Monte-Carlo scenes per value of the varied condition in terms of the \ac{rmspe} \cite{merkofer2024damusic}; all estimators know $D$. The \ac{cnm} is trained with Adam at a learning rate of $10^{-3}$ on batches of 64 scenes simulated on the fly, and the checkpoint with the lowest validation \ac{rmspe} is selected (1024 held-out scenes every 500 steps).

\vspace{-2mm}
\subsection{Baselines}
\vspace{-2mm}

We compare to the classical algorithms \ac{music} \cite{Schmidt1986MUSIC}, Root-\ac{music} \cite{barabell1983rootmusic}, the \ac{mvdr} beamformer \cite{capon1969mvdr}, spatially smoothed \ac{music} (SS-\ac{music}) \cite{shan1985smoothing}, and the deterministic \ac{mle} computed by alternating projection \cite{ziskind1988ml}, all on the nominal manifold, and the stochastic \ac{crb} evaluated with the true manifold \cite{stoica1990crb}. The learned reference algorithms are DA-\ac{music} \cite{merkofer2022damusicicassp, merkofer2024damusic} and the \ac{cnn} of \cite{papageorgiou2021gridcnn} (GridCNN).
Each learned baseline keeps its original objective and trains on the same simulated data as the \ac{cnm}, once per scenario over the full range of the varied condition, whereas the original works train at a single operating point or over a narrow low-\ac{snr} band \cite{merkofer2024damusic, papageorgiou2021gridcnn}.

\vspace{-1mm}
\section{Results} \label{sec:results}
\vspace{-2mm}

We first examine how the learned manifold changes the estimator under model mismatch and when the parameter space extends beyond what is identifiable from the spatial manifold. We then evaluate the resulting estimation accuracy across the full range of data, source, and model conditions in Fig.~\ref{fig:overview}.

Under array imperfections, the \ac{cnm} corrects the manifold toward the observed array response (Fig.~\ref{fig:qualitative}). On the nominal manifold, the two closest of three sources merge into a single broadened peak and the third remains barely above the noise floor. Using the corrected manifold instead yields three sharp peaks at the true \acp{doa}, for both \ac{music} and \ac{mvdr}, although the \ac{cnm} is trained through the \ac{music} null spectrum only (Fig.~\ref{fig:qualitative}, left). The corresponding steering-vector entries are displaced from their nominal toward their true values (center), while the array-error severity $\rho$ can be decoded linearly from the latent representation $\vect{z}$ (right).

Conditioning also allows the manifold to represent parameters that are not identifiable from the nominal spatial response alone. For a \ac{ula}, the far-field response depends on the product $f\sin\theta$, such that a joint scan over $(\theta,f)$ produces ridges rather than isolated peaks on the nominal manifold (Fig.~\ref{fig:joint_carrier}, left). With access to the temporal structure through $\vect{z}$, the corrected landscape separates these ridges into peaks at the true angle--carrier pairs (Fig.~\ref{fig:joint_carrier}, right). 
In the near field, a joint angle--range scan recovers range closely up to approximately 20 half-wavelengths, with increasing scatter toward the far-field end of the evaluated range (Fig.~\ref{fig:readouts}).

Across the data conditions in the top row of Fig.~\ref{fig:overview}, the \ac{cnm} improves upon the learned benchmarks and, except at high \ac{snr}, upon the nominal subspace methods. It follows the \ac{crb} down to $-10$ dB, where the \ac{mle} is considerably less accurate, and from $T=10$ snapshots onward, where nominal \ac{music} is up to an order of magnitude less accurate, and remains largely unaffected by spatially colored noise. With more snapshots, the \ac{mle} comes closer to the bound, and at high \ac{snr}, where Root-\ac{music} approaches the bound, the \ac{cnm} saturates at approximately $0.25^\circ$.

For the source conditions in the middle row, the \ac{cnm} resolves two sources separated by $2^\circ$, which nominal \ac{music} merges, and is the most accurate estimator for separations between $3.5^\circ$ and $5^\circ$. It remains close to the bound for up to five sources and localizes correlated sources without spatial smoothing up to a correlation of 0.9. For fully coherent sources, the rank-deficient source covariance invalidates the underlying subspace test, and only the \ac{mle}, DA-\ac{music}, GridCNN, and SS-\ac{music} remain operable.

Under the model variations in the bottom row, the \ac{cnm} remains accurate across the trained range of array imperfections, while nominal \ac{music} degrades to $11^\circ$. Its accuracy is almost on par with GridCNN and better than DA-\ac{music}. The \ac{mle} is the most accurate estimator on the calibrated array, but, bound to the nominal manifold, it falls behind the \ac{cnm} by a severity of 0.5 and degrades to about $10^\circ$ at 1.5. In the near field, where the nominal far-field manifold fails throughout, the corrected manifold enables \ac{music} to outperform the learned estimators beyond the closest range. For varying carrier frequency, DA-\ac{music} remains the most accurate \ac{doa}-only estimator, while the \ac{cnm} is the most accurate subspace method up to $f/f_c=0.8$. Finally, \acs{cnm}-\acs{mvdr} follows \acs{cnm}-\acs{music} at the expected offset of the \ac{mvdr} beamformer, showing that the correction transfers with the manifold.

\vspace{-1mm}
\section{Conclusion} \label{sec:conclusion}
\vspace{-2mm}
We presented the \ac{cnm}, a learned, observation-conditioned deformation of the nominal array manifold that augments \ac{music} where its model assumptions are violated, while leaving the subspace estimation and interpretable spectrum of the classical method unchanged.
The \ac{cnm} was shown to restore the resolution of \ac{music} under array imperfections, colored noise, correlated sources, and near-field propagation, to recover signal parameters that the spatial response alone leaves ambiguous, and to transfer directly to other estimators such as the \ac{mvdr} beamformer.

\bibliographystyle{styles/IEEEbib}
\bibliography{refs/refs}

@article{krim1996twodecades,
  author  = {Krim, Hamid and Viberg, Mats},
  title   = {Two Decades of Array Signal Processing Research: The Parametric Approach},
  journal = {IEEE Signal Process. Mag.},
  volume  = {13},
  number  = {4},
  pages   = {67--94},
  year    = {1996}
}

@inproceedings{viberg1997twodecades,
  author    = {Viberg, Mats and Krim, Hamid},
  title     = {Two Decades of Statistical Array Processing},
  booktitle = {Proc. Asilomar Conf. Signals, Syst., Comput.},
  pages     = {775--777},
  year      = {1997}
}

@article{pesavento2023threedecades,
  author  = {Pesavento, Marius and Trinh-Hoang, Minh and Viberg, Mats},
  title   = {Three More Decades in Array Signal Processing Research: An Optimization and Structure Exploitation Perspective},
  journal = {IEEE Signal Process. Mag.},
  volume  = {40},
  number  = {4},
  pages   = {92--106},
  year    = {2023}
}

@article{Schmidt1986MUSIC,
  author  = {Schmidt, Ralph O.},
  title   = {Multiple Emitter Location and Signal Parameter Estimation},
  journal = {IEEE Trans. Antennas Propag.},
  volume  = {34},
  number  = {3},
  pages   = {276--280},
  year    = {1986}
}

@article{capon1969mvdr,
  author  = {Capon, Jack},
  title   = {High-Resolution Frequency-Wavenumber Spectrum Analysis},
  journal = {Proc. IEEE},
  volume  = {57},
  number  = {8},
  pages   = {1408--1418},
  year    = {1969}
}

@article{roy1989esprit,
  author  = {Roy, Richard and Kailath, Thomas},
  title   = {{ESPRIT}---Estimation of Signal Parameters via Rotational Invariance Techniques},
  journal = {IEEE Trans. Acoust., Speech, Signal Process.},
  volume  = {37},
  number  = {7},
  pages   = {984--995},
  year    = {1989}
}

@inproceedings{barabell1983rootmusic,
  author    = {Barabell, Arthur J.},
  title     = {Improving the Resolution Performance of Eigenstructure-Based Direction-Finding Algorithms},
  booktitle = {Proc. IEEE Int. Conf. Acoust., Speech, Signal Process. (ICASSP)},
  pages     = {336--339},
  year      = {1983}
}

@article{shan1985smoothing,
  author  = {Shan, Tie-Jun and Wax, Mati and Kailath, Thomas},
  title   = {On Spatial Smoothing for Direction-of-Arrival Estimation of Coherent Signals},
  journal = {IEEE Trans. Acoust., Speech, Signal Process.},
  volume  = {33},
  number  = {4},
  pages   = {806--811},
  year    = {1985}
}

@article{wang1985coherent,
  author  = {Wang, Hong and Kaveh, Mostafa},
  title   = {Coherent Signal-Subspace Processing for the Detection and Estimation of Angles of Arrival of Multiple Wide-Band Sources},
  journal = {IEEE Trans. Acoust., Speech, Signal Process.},
  volume  = {33},
  number  = {4},
  pages   = {823--831},
  year    = {1985}
}

@article{huang1991nearfield,
  author  = {Huang, Yung-Dar and Barkat, Mourad},
  title   = {Near-Field Multiple Source Localization by Passive Sensor Array},
  journal = {IEEE Trans. Antennas Propag.},
  volume  = {39},
  number  = {7},
  pages   = {968--975},
  year    = {1991}
}

@article{stoica1990crb,
  author  = {Stoica, Petre and Nehorai, Arye},
  title   = {Performance Study of Conditional and Unconditional Direction-of-Arrival Estimation},
  journal = {IEEE Trans. Acoust., Speech, Signal Process.},
  volume  = {38},
  number  = {10},
  pages   = {1783--1795},
  year    = {1990}
}

@article{liu2023sam,
  author  = {Liu, Wei and Haardt, Martin and Greco, Maria S. and Mecklenbr{\"a}uker, Christoph and Willett, Peter},
  title   = {Twenty-Five Years of Sensor Array and Multichannel Signal Processing: A Review of Progress to Date and Potential Research Directions},
  journal = {IEEE Signal Process. Mag.},
  volume  = {40},
  number  = {4},
  pages   = {80--91},
  year    = {2023}
}

@article{ziskind1988ml,
  author  = {Ziskind, I. and Wax, M.},
  title   = {Maximum Likelihood Localization of Multiple Sources by Alternating Projection},
  journal = {IEEE Trans. Acoust., Speech, Signal Process.},
  volume  = {36},
  number  = {10},
  pages   = {1553--1560},
  year    = {1988},
}

@article{swindlehurst1992modelerrors,
  author  = {Swindlehurst, A. Lee and Kailath, Thomas},
  title   = {A Performance Analysis of Subspace-Based Methods in the Presence of Model Errors, Part {I}: The {MUSIC} Algorithm},
  journal = {IEEE Trans. Signal Process.},
  volume  = {40},
  number  = {7},
  pages   = {1758--1774},
  year    = {1992}
}

@article{friedlander1991coupling,
  author  = {Friedlander, Benjamin and Weiss, Anthony J.},
  title   = {Direction Finding in the Presence of Mutual Coupling},
  journal = {IEEE Trans. Antennas Propag.},
  volume  = {39},
  number  = {3},
  pages   = {273--284},
  year    = {1991}
}

@article{vorobyov2003robust,
  author  = {Vorobyov, Sergiy A. and Gershman, Alex B. and Luo, Zhi-Quan},
  title   = {Robust Adaptive Beamforming Using Worst-Case Performance Optimization: A Solution to the Signal Mismatch Problem},
  journal = {IEEE Trans. Signal Process.},
  volume  = {51},
  number  = {2},
  pages   = {313--324},
  year    = {2003}
}

@article{belloni2007manifold,
  author  = {Belloni, Fabio and Richter, Andreas and Koivunen, Visa},
  title   = {{DoA} Estimation via Manifold Separation for Arbitrary Array Structures},
  journal = {IEEE Trans. Signal Process.},
  volume  = {55},
  number  = {10},
  pages   = {4800--4810},
  year    = {2007}
}

@article{liu2018imperfections,
  author  = {Liu, Zhang-Meng and Zhang, Chenwei and Yu, Philip S.},
  title   = {Direction-of-Arrival Estimation Based on Deep Neural Networks with Robustness to Array Imperfections},
  journal = {IEEE Trans. Antennas Propag.},
  volume  = {66},
  number  = {12},
  pages   = {7315--7327},
  year    = {2018}
}

@article{papageorgiou2021gridcnn,
  author  = {Papageorgiou, Georgios K. and Sellathurai, Mathini and Eldar, Yonina C.},
  title   = {Deep Networks for Direction-of-Arrival Estimation in Low {SNR}},
  journal = {IEEE Trans. Signal Process.},
  volume  = {69},
  pages   = {3714--3729},
  year    = {2021}
}

@article{barthelme2021reconstruction,
  author  = {Barthelme, Andreas and Utschick, Wolfgang},
  title   = {{DoA} Estimation Using Neural Network-Based Covariance Matrix Reconstruction},
  journal = {IEEE Signal Process. Lett.},
  volume  = {28},
  pages   = {783--787},
  year    = {2021}
}

@inproceedings{merkofer2022damusicicassp,
  author    = {Merkofer, Julian P. and Revach, Guy and Shlezinger, Nir and van Sloun, Ruud J. G.},
  title     = {Deep Augmented {MUSIC} Algorithm for Data-Driven {DoA} Estimation},
  booktitle = {Proc. IEEE Int. Conf. Acoust., Speech, Signal Process. (ICASSP)},
  pages     = {3598--3602},
  year      = {2022}
}

@article{merkofer2024damusic,
  author  = {Merkofer, Julian P. and Revach, Guy and Shlezinger, Nir and Routtenberg, Tirza and van Sloun, Ruud J. G.},
  title   = {{DA-MUSIC}: Data-Driven {DoA} Estimation via Deep Augmented {MUSIC} Algorithm},
  journal = {IEEE Trans. Veh. Technol.},
  volume  = {73},
  number  = {2},
  pages   = {2771--2785},
  year    = {2024}
}

@inproceedings{shmuel2023deeprootmusic,
  author    = {Shmuel, Dor H. and Merkofer, Julian P. and Revach, Guy and van Sloun, Ruud J. G. and Shlezinger, Nir},
  title     = {Deep Root {MUSIC} Algorithm for Data-Driven {DoA} Estimation},
  booktitle = {Proc. IEEE Int. Conf. Acoust., Speech, Signal Process. (ICASSP)},
  pages     = {1--5},
  year      = {2023}
}

@article{shmuel2025subspacenet,
  author  = {Shmuel, Dor H. and Merkofer, Julian P. and Revach, Guy and van Sloun, Ruud J. G. and Shlezinger, Nir},
  title   = {{SubspaceNet}: Deep Learning-Aided Subspace Methods for {DoA} Estimation},
  journal = {IEEE Trans. Veh. Technol.},
  volume  = {74},
  number  = {3},
  pages   = {4962--4976},
  year    = {2025}
}

@article{shlezinger2023modelbased,
  author  = {Shlezinger, Nir and Whang, Jay and Eldar, Yonina C. and Dimakis, Alexandros G.},
  title   = {Model-Based Deep Learning},
  journal = {Proc. IEEE},
  volume  = {111},
  number  = {5},
  pages   = {465--499},
  year    = {2023}
}

@inproceedings{lee2019settransformer,
  author    = {Lee, Juho and Lee, Yoonho and Kim, Jungtaek and Kosiorek, Adam R. and Choi, Seungjin and Teh, Yee Whye},
  title     = {Set Transformer: A Framework for Attention-Based Permutation-Invariant Neural Networks},
  booktitle = {Proc. Int. Conf. Mach. Learn. (ICML)},
  pages     = {3744--3753},
  year      = {2019}
}

@article{xie2022neuralfields,
  author  = {Xie, Yiheng and Takikawa, Towaki and Saito, Shunsuke and Litany, Or and Yan, Shiqin and Khan, Numair and Tombari, Federico and Tompkin, James and Sitzmann, Vincent and Sridhar, Srinath},
  title   = {Neural Fields in Visual Computing and Beyond},
  journal = {Comput. Graph. Forum},
  volume  = {41},
  number  = {2},
  pages   = {641--676},
  year    = {2022}
}

@article{yue2026sphereencoder,
  author  = {Yue, Kaiyu and Jia, Menglin and Hou, Ji and Goldstein, Tom},
  title   = {Image Generation with a Sphere Encoder},
  journal = {arXiv preprint arXiv:2602.15030},
  year    = {2026}
}

\end{document}